\documentclass[%
  reprint,
  amsmath,
  amssymb,
  aps,
  pra
]{revtex4-2}

\newcommand\paperTitle{Bosonic quantum dynamics following colliding potential wells}

\usepackage{amsmath}
\usepackage{amssymb}
\usepackage{braket}
\usepackage{hyperref}
\usepackage{xcolor}
\usepackage{xfrac}
\usepackage{xparse}
\definecolor{correction}{HTML}{114076}

\hypersetup{%
  colorlinks,
  unicode=true,
  pdftitle={\paperTitle},
  pdfauthor={Fabian Köhler},
  pdfcreator={Fabian Köhler},
  pdfproducer={Fabian Köhler},
}

\DeclareMathOperator{\Tr}{Tr}

\newcommand{\ev}[1]{\langle{#1}\rangle}

\begin{document}
\title{\paperTitle}
\author{Fabian K\"ohler}
\email{fkoehler@physnet.uni-hamburg.de}
\affiliation{Center for Optical Quantum Technologies, Department of Physics, University of Hamburg, Luruper Chaussee 149, 22761 Hamburg Germany}
\author{Peter Schmelcher}
\email{pschmelc@physnet.uni-hamburg.de}
\affiliation{Center for Optical Quantum Technologies, Department of Physics, University of Hamburg, Luruper Chaussee 149, 22761 Hamburg Germany}
\affiliation{The Hamburg Centre for Ultrafast Imaging, University of Hamburg, Luruper Chaussee 149, 22761 Hamburg, Germany}
\date{\today}

\begin{abstract}
  We employ the multiconfiguration time-dependent Hartree method for bosons in order to investigate the correlated nonequilibrium quantum dynamics of two bosons confined in two colliding and uniformly accelerated Gaussian wells.
  As the wells approach each other an effective, transient double-well structure is formed.
  This induces a transient and oscillatory over-barrier transport.
  We monitor both the amplitude of the intra well dipole mode in the course of the dynamics as well as the final distribution of the particles between the two wells.
  For fast collisions we observe an emission process which we attribute to two distinct mechanisms.
  Energy transfer processes lead to an untrapped fraction of bosons and a resonant enhancement of the deconfinement for certain kinematic configurations can be observed.
  Despite the comparatively weak interaction strengths employed in this work, we identify strong interparticle correlations by analyzing the corresponding Von Neumann entropy.
\end{abstract}

\maketitle

\section{Introduction}

Ever since the first realizations of Bose-Einstein condensates~\cite{davis1995,bradley1995,anderson1995}, ultracold quantum gases were the focus of experimental and theoretical research in quantum physics.
Their nearly perfect isolation from the environment as well as their excellent tunability render them ideal platforms to simulate a wide variety of quantum many-body systems~\cite{bloch2008,polkovnikov2011,bloch2012} in order to unravel their fundamental physical properties.
Experimental advancements in recent years have enabled the study of ensembles of ultracold atoms with a controlled number of particles~\cite{serwane2011,kaufman2014} confined in almost arbitrarily shaped external potentials~\cite{henderson2009} like optical lattices~\cite{jaksch1998,bloch2005}, harmonic traps~\cite{chu1986}, and ring traps~\cite{morizot2006}.
By varying the confinement it is possible to realize effectively three-dimensional~\cite{greiner2002a,duan2003}, two-dimensional~\cite{zobay2001,colombe2004}, and one-dimensional~\cite{orzel2001,paredes2004} systems.
Magnetic Feshbach~\cite{kohler2006,chin2010} and confinement-induced resonances~\cite{olshanii1998,kim2006,giannakeas2012,giannakeas2013} provide fine-grained control of the interparticle interaction.
Recent studies have employed this versatile toolbox of ultracold atoms to establish links to solid-state systems~\cite{anderson1998,jo2009}, the electronic structure of molecules~\cite{luhmann2015}, light-matter interaction~\cite{sala2017}, topological matter~\cite{jotzu2014,goldman2016}, and even black-hole analogs~\cite{steinhauer2016}.

In recent years, optical tweezers have become important instruments to confine and move microscopic objects by exerting small forces via highly focused laser beams.
This tool was originally developed to manipulate micrometer-sized particles~\cite{ashkin1970,ashkin1986} but was later refined to manipulate objects on many different length scales ranging from individual atoms~\cite{gaetan2009,saskin2019} to bacteria and viruses~\cite{ashkin1987}.
These advancements sparked strong interest in using optical tweezers for the precise manipulation of ensembles of ultracold neutral atoms~\cite{roberts2014} including Rydberg atoms~\cite{schymik2020,scholl2020,morgado2020}.
A very interesting direction of research is to use multiple optical tweezers to accelerate atomic clouds~\cite{rakonjac2012}, which allows one to set up optical colliders~\cite{kjaergaard2004,thomas2017,thomas2018}.
In these experiments, fundamental properties of quantum scattering processes were observed such as partial wave interference or the loss of particles in resonant collisions.
In this light, colliding ultracold atoms could be used to mimic electrons during atom-atom collisions.
Since the dynamics of ultracold atoms takes place on much larger time scales, the usually very fast electronic processes could be slowed down~\cite{rajagopal2017,sala2017,senaratne2018}, potentially providing in depth insights into the fundamental processes of atom-atom or atom-ion collisions such as projectile ionization~\cite{wang1993,montenegro1994} or charge transfer~\cite{fite1960,olson1977}.

Another interesting application of ultracold atoms is quantum information processing~\cite{bloch2008a}.
In this context, time-dependent colliding trap potentials have been proposed for the realization of two-qubit quantum gates as well as the efficient creation of highly entangled states~\cite{jaksch1999,calarco2000}, which are two essential features required for a quantum computer.

In the present investigation two bosonic particles are confined in two colliding Gaussian potential wells.
We solve this time-dependent problem using the ab initio multiconfiguration time-dependent Hartree method for bosons (MCTDHB), which provides an exact description capturing all correlations~\cite{alon2008,cao2017}.
This allows us to compute the time evolution of the two-body wave function across a wide range of kinematic parameters in contrast to the other theoretical investigations of colliding potentials in the literature~\cite{jaksch1999,calarco2000} which relied on employing effective models and were limited to adiabatic movements of the traps.
We show that during the time evolution of this system an effective time-dependent double-well structure forms that drives an oscillatory over-barrier bosonic transport between the wells.
This process terminates when the wells have been separated sufficiently after penetrating each other.
During the collision process the displacement of the bosons from the well trajectories induces an intra well dipole mode and determines the final distribution of the particles between the wells.
For fast collisions this setup exhibits deconfinement of the particles, which we can attribute to two different mechanisms.
First, for very fast accelerations an increase in kinetic energy leads to a positive total energy of the system towards the end of the time evolution thereby causing an untrapping of particles.
Second, we observe a resonant enhancement of the emission for certain kinematic parameters similar to the ionization processes that take place in atom-atom collisions.

Our work is structured as follows.
In Sec.~\ref{sec:setup} we introduce the physical setup and describe the computational approach used to solve the time-dependent problem.
We proceed by presenting the results for the dynamics of two interacting bosonic particles in Sec.~\ref{sec:results} and discuss suitable observables to unravel the properties of the system.
We summarize our findings in Sec.~\ref{sec:conclusions} and provide an outlook on possible future studies.
Finally, we comment on the convergence of our variational MCTDHB approach in the Appendix.

\section{Physical setup and computational approach}\label{sec:setup}
In the present work we investigate the nonequilibrium quantum dynamics of a closed system of $N=2$ interacting bosons.
We employ the MCTDHB~\cite{alon2007,streltsov2007,alon2008} to solve the time-dependent many-body {Schr\"odinger equation and gain access to the correlated quantum dynamics of the particles.
This approach employs a time-dependent, variationally optimal basis ${\lbrace \phi_i(x,t)\rbrace}_{i=1}^{M}$ of $M$ single-particle functions (SPFs).
The many-body wave function $\ket{\Psi(t)}$ is then expanded as a superposition
\begin{equation}
  \label{eq:wave_function_ansatz}
  \ket{\Psi(t)}=\sum\limits_{\vec{n}|N}C_{\vec{n}}(t)\ket{\vec{n};t}
\end{equation}
of all $\binom{N+M-1}{N}$ time-dependent $N$-particle number states $\ket{\vec{n};t}$ that can be built from the $M$ SPFs using time-dependent coefficients $C_{\vec{n}}(t)$.
Finally, the Lagrangian formulation of the time-dependent variational principle~\cite{kramer1981,broeckhove1988} yields equations of motions for the SPFs and the coefficients~\cite{alon2007,alon2008} are then solved numerically.
The MCTDHB provides access to the time evolution of the full many-body wave function, which allows us to compute all relevant characteristics of the underlying system.

We consider $N=2$ bosons of mass $m$ interacting repulsively with a contact interaction of strength of $g$~\cite{pitaevskii2003,pethick2008}.
The Hamiltonian of the system reads
\begin{equation}
  \label{eq:many_body_hamiltonian}
  H(\lbrace x_i\rbrace, t)=\sum\limits_{i=1}^N h(x_i,t)+g\sum\limits_{\substack{i,j=1\\i<j}}^{N}\delta(x_i-x_j).
\end{equation}
The one-body Hamiltonian
\begin{equation}
  \label{eq:one_body_hamiltonian}
  h(x,t)=-\frac{\hbar^2}{2m}\frac{\partial^2}{\partial x^2}+V(x,t)
\end{equation}
acts on each particle individually and includes both a kinetic term and the external potential $V(x,t)$.

In our setup, the external potential
\begin{equation}
  \label{eq:potential}
  V(x,t)=-V_0\exp\left(-{\left(\frac{x-\mu(t)}{\sqrt{2}\sigma}\right)}^2\right)-V_0^\prime\exp\left(-{\left(\frac{x-\mu^\prime(t)}{\sqrt{2}\alpha\sigma}\right)}^2\right)
\end{equation}
consists of two Gaussian wells of depths $V_0$ and $V_0^\prime$ centered around $\mu(t)$ and $\mu^\prime(t)$, which approach each other in the first phase of the collision process (see Fig.~\ref{fig:setup}).
The width of the two Gaussians is characterized by their standard deviations $\sigma$ and $\alpha\sigma$, where $\alpha$ is a dimensionless asymmetry factor.
We drive the nonequilibrium dynamics by a motion of the well centers specified by the expectation values $\mu(t)$ and $\mu^\prime(t)$.
Hence, the potential~\eqref{eq:potential} and consequently the Hamiltonians~\eqref{eq:one_body_hamiltonian} and~\eqref{eq:many_body_hamiltonian} are time dependent.

The investigation of the physical system can be greatly simplified by employing a suitable unit system.
We rescale all positions using the length unit $l_{\mathrm{G}}=\sqrt{2}\sigma$ and all energies using the energy unit $E_{\mathrm{G}}=\hbar^2{(2m\sigma^2)}^{-1}$ in order to obtain a dimensionless formulation and to eliminate both $\sigma$ and $m$ as physical parameters from the potential and Hamiltonian.
The corresponding time unit $t_{\mathrm{G}}=2m\sigma^2\hbar^{-1}$ can be inferred from the Schr\"odinger equation.
For the analysis of the dynamics it is instructive to additionally introduce the unit $v_{\mathrm{G}}=\hbar{(\sqrt{2}m\sigma)}^{-1}$ for speeds.

The dynamics of the particles strongly depends on the initial state.
A natural choice is to prepare the system in the ground state of the initial many-body Hamiltonian $H(\lbrace x_i\rbrace, t=0)$ where the particles would be delocalized over the two wells.
However, we will use the ground state for $V_0^\prime =0$ which results in all particles being located in the left well centered around $\mu(0)$ (see Fig.~\ref{fig:setup}).
This allows us to track them during the transport processes that occur during the time propagation.
This initial state can be computed efficiently using the improved relaxation algorithm~\cite{meyer2003}.
Experimentally, such a state could be prepared with high fidelity by loading two atoms in a single optical microtrap and then slowly ramping on the spatially separated potential wells~\cite{serwane2011,lester2015,murmann2015}.
\begin{figure}[htbp]
  \centering
  \includegraphics{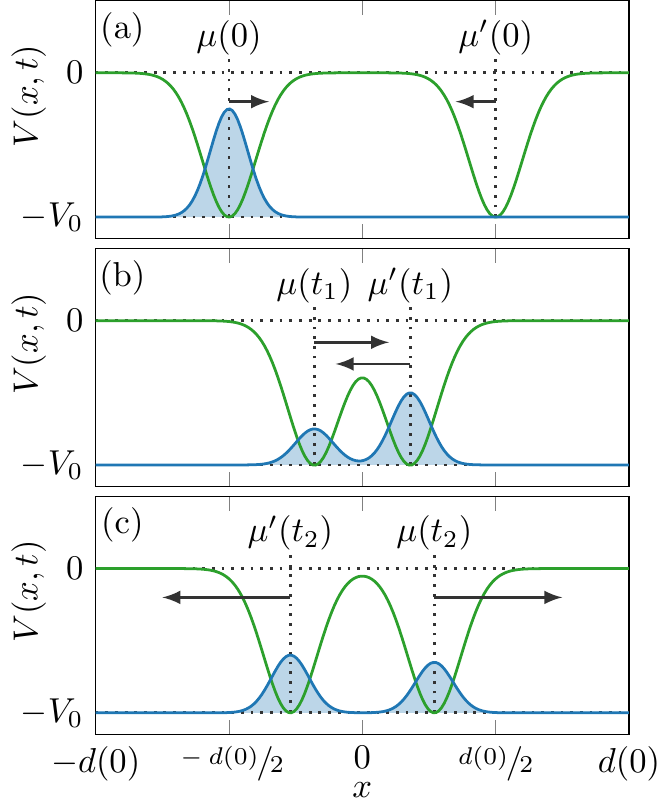}
  \caption{Sketch of the system at different points in time $t_0=0<t_1<t_2$ during the dynamics.
    The green line indicates the external trapping potential consisting of two Gaussian wells while the blue line symbolizes the spatial distribution of the particles.
    (a) The time-evolution of the system starts with the interacting ground state in the left well.
    (b) As the wells accelerate towards each other, a transient, time-dependent double-well structure forms.
    (c) After the wells penetrated each other they separate again moving in opposite directions.}\label{fig:setup}
\end{figure}

We assume that for $t=0$ the potential wells are at rest.
The most evident choice for the trajectory of the Gaussian well centers $\mu(t)$ and $\mu^\prime(t)$ would be a uniform motion, i.e.\ by boosting the wells to fixed speeds instantaneously.
However, this approach would pump a great deal of energy into the system thereby causing major excitations which would render the dynamics very ``irregular.''
Therefore, we choose to accelerate the wells uniformly towards each other using parabolic trajectories
\begin{align}
  \mu(t)=\mu(0)+\frac{1}{2}at^2 \\
  \mu^{\prime}(t)=\mu^{\prime}(0)-\frac{1}{2}at^2
\end{align}
for the well centers.
Initially, the wells are located symmetrically around $x=0$, i.e., $\mu(0)=-\mu^\prime(0)$ with a separation of $d(0)$.
The propagation is terminated at the final time
\begin{equation}
  \label{eq:final_time}
  t_{\mathrm{f}}=\sqrt{2\frac{d(0)}{a}}
\end{equation}
when the wells have moved through each other and reached their initial separation again.
At this point in time the wells have reached their final speed of $v_{\mathrm{f}}=at_{\mathrm{f}}=\sqrt{2ad(0)}$.

\section{Discussion of the collisional dynamics}\label{sec:results}
In the scope of the present work we limit ourselves to $N=2$ particles when investigating the setup described in Sec.~\ref{sec:setup} in order to unravel the main signatures of the dynamics of the system.
This provides an ideal starting point for future works addressing the case of larger particle numbers.
We choose wells of equal width, i.e., $\alpha=1$, and depth $V_0=V_0^\prime=20E_{\mathrm{G}}$, which are deep enough to support ten trapped states of the one-body Hamiltonian~\eqref{eq:one_body_hamiltonian}.
Initially, the wells are located at $\mu(0)=-3.5l_{\mathrm{G}}$ and $\mu^\prime(0)=3.5l_{\mathrm{G}}$, which corresponds to an initial separation of $d(0)=7l_{\mathrm{G}}$.
For the interaction strength we choose a value of $g=0.5E_{\mathrm{G}}l_{\mathrm{G}}$, which is comparable to an interaction strength of $g_{\mathrm{HO}}\approx 0.199$ in harmonic-oscillator units.
We find that for this value of $g$, $M=6$ SPFs are sufficient for the convergence of our MCTDHB simulations (see Sec.~\ref{sec:convergence}).
We solve the time-dependent problem for varying values of the acceleration $a$ chosen such that the corresponding inverse final speeds $v_{\mathrm{f}}^{-1}$ are equally spaced in the interval $\left[0.1 v_{\mathrm{G}}^{-1},2.5 v_{\mathrm{G}}^{-1}\right]$.
The reason for this choice will become apparent during the analysis since many quantities scale with the inverse speed.

\subsection{Time evolution of the one-body density}
In order to analyze the dynamics of the system and to guide our further analysis approach, we inspect the one-body density~\cite{sakmann2008,dirac1930}
\begin{equation}
  \label{eq:gpop}
  \rho^{(1)}(x,t)=N\int{\lvert\Psi(x,x_2,\ldots,x_N,t)\rvert}^2 dx_2\ldots dx_N,
\end{equation}
with $N=2$ in our case.
This quantity provides insight into the temporal evolution of the spatial distribution of the particles since $\rho^{(1)}(x,t)$ corresponds to the probability density of finding a particle at the position $x$ at the time $t$.

Figures 2(a)--2(f) show the time evolution of $\rho^{(1)}(x,t)$ for various values of the acceleration which correspond to different inverse final speeds $v_{\mathrm{f}}^{-1}$.
If the acceleration is not too fast [see Figs. 2(a)--2(e)], we can identify three distinct stages of the dynamics indicated by (I)--(III).
\begin{figure}[htbp]
  \centering
  \includegraphics{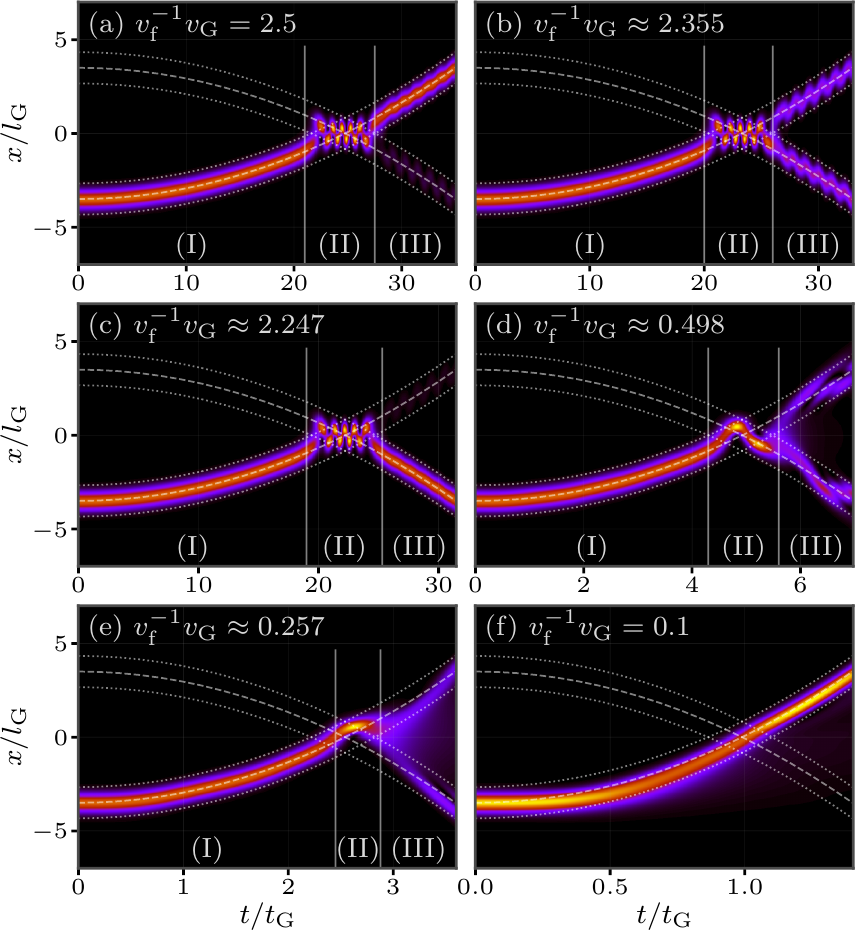}
  \caption{Time evolution of the one-body density $\rho^{(1)}(x,t)$ [see Eq.~\eqref{eq:gpop}] for different inverse final speeds $v_{\mathrm{f}}^{-1}\propto a^{-1/2}$. The dashed white lines indicate the trajectories of the well centers while the dotted white lines indicate the positions of the FWHM of the Gaussian wells\label{fig:gpop}.}
\end{figure}

The particles are initially localized in the well centered at $\mu(0)=-3.5l_{\mathrm{G}}$ and follow its parabolic trajectory $\mu(t)$ during stage (I) of the dynamics while wells approach each other.
No effect of the presence of the second well centered around $\mu^\prime(t)$ is visible during this phase of the dynamics.
During stage (II) the wells are in close proximity and they even penetrate each other.
Hence, an effective double-well structure forms (see Fig.~\ref{fig:setup}) that changes its shape over time and we observe a collective oscillatory particle transport over the central barrier from the left to the right well and vice versa.
Towards the end of the propagation, during stage (III), we find several effects depending on the acceleration and hence $v_{\mathrm{f}}^{-1}$.
In general, the particles are delocalized over both wells with varying ratios.
For certain values of $v_{\mathrm{f}}^{-1}$ however, the bosons are almost completely localized in one of the wells.
Additionally, we observe a sloshing motion of the particles within each well.
We characterize this motion as a dipole mode~\cite{pitaevskii2003,pethick2008} since the center-of-mass (c.m.) position of the particles oscillates around the center of the wells in which they are confined.
This collective excitation is accompanied by a breathing mode which manifests in a periodic widening and contraction of the atomic cloud in each well.
However, the breathing is much less pronounced compared to the dipole oscillation such that we refer to the sloshing motion as a dipole mode in the following.
Generally, we observe that the one-body density is well contained within one full width at half maximum (FWHM) around the well centers as indicated by the white lines in Fig.~\ref{fig:gpop}.
However, for fast collisions [see Fig.~\ref{fig:gpop} (e)] we notice a faint density halo in the region between the wells, which indicates an untrapped fraction of particles, i.e., a finite probability of detecting a particle in this region.
When moving towards even faster accelerations we also observe effects of the inertia of the bosons [see Fig.~\ref{fig:gpop} (f)] which seem to move more slowly than the left well and leave the FWHM region before finally catching up with the well towards the end of the dynamics.

\subsection{Center of mass position}
In order to analyze the transport of particles, we introduce the c.m.\ position
\begin{equation}
  \label{eq:com_operator}
  \ev{X}(t)=\frac{1}{N}\sum\limits_{i=1}^{N}\ev{x_i}(t)
\end{equation}
which measures the average position of the particles.
In Figs.~3(a) and 3(b) we show two examples for the time evolution of this quantity.
We can clearly make out the three aforementioned phases (I)--(III) of the dynamics.
During stage (I) of the time evolution, $\ev{X}(t)$ matches the trajectory of the left well $\mu(t)$ as the particles simply follow the motion of the potential.
In part (II) we observe an oscillation of $\ev{X}(t)$ around $0$ which indicates the oscillatory particle transport in the effective double-well structure from the left to the right well and vice versa.
During stage (III) we notice that the evolution of $\ev{X}(t)$ strongly depends on the kinematic parameters.
For some values of $v_{\mathrm{f}}^{-1}v_{\mathrm{G}}$, $\ev{X}(t)$ closely follows one of the trajectories $\mu(t)$ and $\mu^\prime(t)$ and the dipole mode vanishes [see Fig.~\ref{fig:com} (b)].
In other cases [see Fig.~\ref{fig:com} (a)] $\ev{X}(t)$ lies in the region between $\mu(t)$ and $\mu^\prime(t)$ and the dipole mode is well pronounced.
The amplitude of the dipole mode varies depending on $a$ and is maximal when $\ev{X}(t)$ oscillates close to zero.

As the next step, we quantify the number of transport processes during phase (II) of the dynamics by determining the number of zero crossings $N_{\mathrm{ZC}}^{\mathrm{(II)}}$ of the signal $\ev{X}(t)$ for each value of $v_{\mathrm{f}}^{-1}$ during this stage [see Fig.~\ref{fig:com} (d)].
Here $N_{\mathrm{ZC}}^{\mathrm{(II)}}$ increases monotonically with $v_{\mathrm{f}}^{-1}$ since the effective double-well structure persists for a longer time period and more oscillations can take place.
Since the number of zero crossings has to be a non-negative integer, $N_{\mathrm{ZC}}^{\mathrm{(II)}}$ is a step function of $v_{\mathrm{f}}^{-1}$.
We find the step width to be approximately equal for all steps with an average width of $0.221v_{\mathrm{G}}^{-1}$.
\begin{figure}[htbp]
  \centering
  \includegraphics{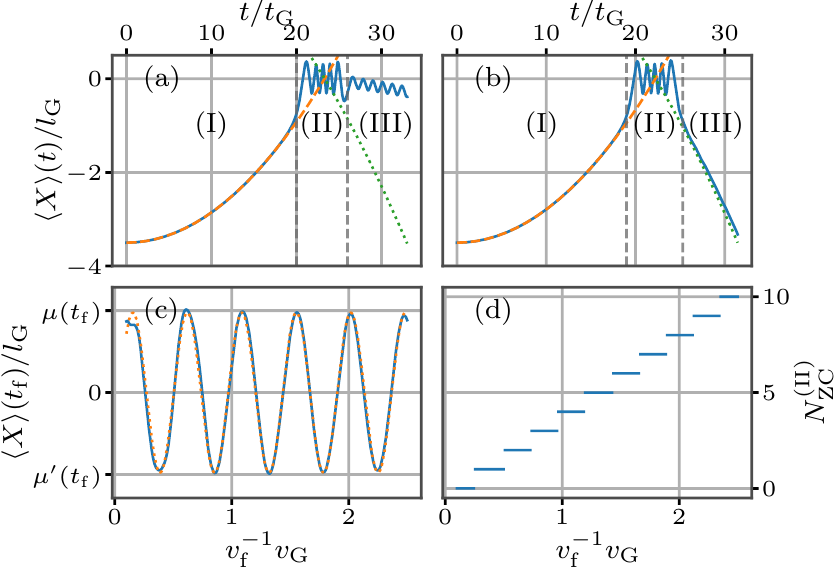}
  \caption{Time-evolution of the c.m.\ position (blue solid line) as a function of time for (a) $v_{\mathrm{f}}^{-1}v_{\mathrm{G}}\approx 2.355$ and (b) $v_{\mathrm{f}}^{-1}v_{\mathrm{G}}\approx 2.247$. The orange dashed line indicates the trajectory $\mu(t)$ while the green dotted line visualizes $\mu^\prime(t)$. (c) Expectation value of the c.m.\ position of the particles in the final state as a function of $v_{\mathrm{f}}^{-1}$. The orange dashed line corresponds to a cosine fit of the signal. (d) Number of zero crossings $N_{\mathrm{ZC}}^{\mathrm{(II)}}$ of $\ev{X}(t)$ in the region (II) as a function of $v_{\mathrm{f}}^{-1}$.\label{fig:com}}
\end{figure}

As mentioned before, the final location of the particles strongly depends on the acceleration $a$.
Figure~\ref{fig:com} (c) shows the final c.m.\ position of the particles $\ev{X}(t_{\mathrm{f}})$ as a function of $v_{\mathrm{f}}^{-1}$, which resembles a cosine like structure.
Using a least-squares fit, we can extract the period $\Delta v^{-1}=0.47v_{\mathrm{G}}^{-1}$ and the amplitude $3.42l_{\mathrm{G}}$ of the signal.
From the amplitude of the oscillation, we can deduce that indeed for certain values of $v_{\mathrm{f}}^{-1}$ the density is almost completely located in one of the wells.
A value of $\ev{X}(t_{\mathrm{f}})=\pm 3.5l_{\mathrm{G}}$ would indicate that the average position of the particles coincides with the final position of one of the well centers.
For most values of $v_{\mathrm{f}}^{-1}$ however, the final center-of-mass position lies somewhere between these extreme cases and indicates that the particles are delocalized across both wells.

A further analysis of the center-of-mass motion shows that the final distribution of the particles as well as the amplitude of the dipole mode depend on the displacement of the c.m.\ position from the trajectories of the wells at the transition from stage (II) to (III) of the dynamics.
If the c.m.\ position $\ev{X}(t)$ is close to one of the well centers at this transition point, the particles get pinned in that particular well.
A small deflection of $\ev{X}(t_{\mathrm{f}})$ from the well center leads then to small amplitudes of the corresponding dipole mode in this well.
For most values of $v_{\mathrm{f}}^{-1}$ however, the separation of the wells splits the one-body density into two parts and the particles are delocalized across both wells.
As emphasized, the displacement of the particles within the wells induces an intra-well dipole mode, the amplitude of which is maximal if $\ev{X}(t)$ is close to $0$ at the transition from stage (II) to (III), which corresponds to the maximal deflection of the particles from the well center.

\begin{figure}[htbp]
  \centering
  \includegraphics{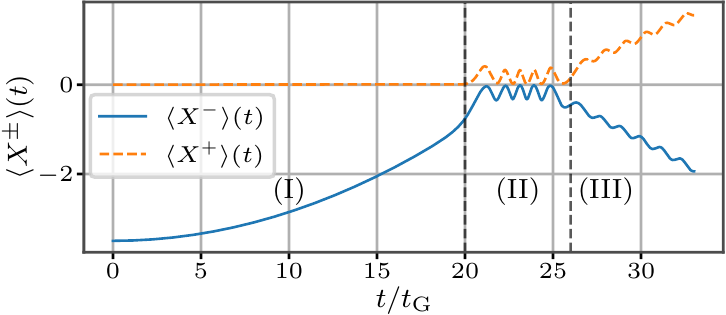}
  \caption{Time-evolution of the truncated c.m.\ observables $\ev{X^{\pm}}(t)$ [see Eq.\ \ref{eq:comTruncated}] for $v_{\mathrm{f}}^{-1}v_{\mathrm{G}}\approx 2.355$.\label{fig:truncatedCom}}
\end{figure}
In order to distinguish between the intra well dynamics different wells, we introduce the truncated c.m.\ observables
\begin{equation}
  \ev{X^{\pm}}(t)=\frac{1}{N}\sum\limits_{i=1}^{N}\ev{x_i\Theta(\pm x_i)}(t)\label{eq:comTruncated}.
\end{equation}
which measure the average position of particles on either the positive or the negative side with respect to $x=0$.
Figure~\ref{fig:truncatedCom} shows an example for the time evolution of these observables.
Here $\ev{X^{+}}(t)$ is zero during phase (I) of the dynamics as the particles are initially contained in the left well and follow its trajectory.
The periodic transport in the transient double-well potential during phase (II) is clearly visible.
During part (III) of the dynamics, the dipole motion of the particles in the initially left (right) well manifests itself in an oscillatory modulation of $\ev{X^{+}}(t)$ [$\ev{X^{-}}(t)$].
By analyzing the turning points of these modulations, we determine a phase of $\sfrac{\pi}{2}$ between the two oscillations.
Furthermore, we notice that the oscillation period of both observables lies in the range $0.55t_{\mathrm{G}}- 0.6t_{\mathrm{G}}$ and is approximately constant across all values of $a$ which is to be expected since the frequency of the dipole mode only depends on the shape of the potential well.

\subsection{Nature of the particle transport}\label{sec:transport}
In order to classify the transport process between the left and right well that takes place in phase (II) of the dynamics, we analyze the two-body wave function $\ket{\Psi(t)}$ with respect to the time-dependent one-body Hamiltonian $h(x,t)$ (Equation~\eqref{eq:one_body_hamiltonian}).
We consider the instantaneous eigenbasis of $h(x,t)$ spanned by the time-dependent eigenstates $\lbrace\ket{\Phi_i(t)} \rbrace$ with the corresponding eigenenergies $\varepsilon_i(t)$, i.e.\ $h(x,t)\Phi_i(x,t)=\varepsilon_i(t)\Phi_i(x,t)$, while assuming an energetic ordering $\varepsilon_i(t)\leq\varepsilon_{i+1}(t)$ for all times.
Figure~\ref{fig:oneBodySpectrum} shows the eigenenergies of the ten energetically lowest eigenstates as a function of the well separation $d(t)=d(0)-at^2$.
At the initial [$d(0)$] and final [$d(t_{\mathrm{f}})$] separations, the external potential is able to support ten trapped eigenstates, i.e., states with negative eigenenergies, which are pairwise degenerate.
It should be noted that for positive energies the system exhibits a discrete spectrum of untrapped states instead of a continuous spectrum of extended continuum states since we employ a finite grid for the numerical treatment of the problem which imposes periodic boundary conditions (see the Appendix).
However, this does not impact our analysis of the trapped fraction or the occupation of the trapped states.
If the wells reach close proximity, an effective double-well structure forms (see Fig.~\ref{fig:setup}), where $V(x=0)$ determines the height of the barrier and the energetic degeneracies are lifted.
In the vicinity of $d(t)=0$ the central barrier vanishes and the external potential is a single Gaussian well centered around $x=0$ with a depth $V(x=0)=-2V_0$.
Here the eigenenergies $\varepsilon_7(t)$, $\varepsilon_8(t)$ and $\varepsilon_9(t)$ cross zero and reach positive value such that the associated eigenstates become untrapped.
\begin{figure}[htpb]
  \centering
  \includegraphics{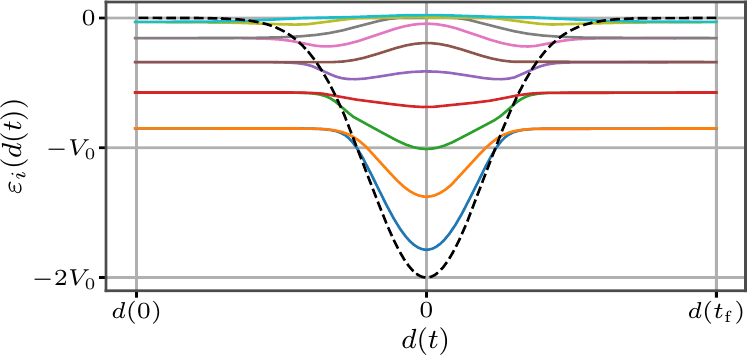}
  \caption{Spectrum of the one-body Hamiltonian $h(x,t)$ [\eqref{eq:one_body_hamiltonian}] as a function of the well separation $d(t)$. We show the ten energetically lowest eigenenergies (colored solid lines) and the values of the central potential $V(x=0)$ (black dashed line).\label{fig:oneBodySpectrum}}
\end{figure}

We proceed with our analysis by defining the operator
\begin{equation}
  P_j(t)=\frac{1}{N}\sum\limits_{i=1}^{N}\ket{\Phi_j^i(t)}\bra{\Phi_j^i(t)}
\end{equation}
where $\ket{\Phi_j^i(t)}\bra{\Phi_j^i(t)}$ projects the $i$th particle onto the $j$th one-body eigenstate $\ket{\Phi_j(t)}$.
Computing the expectation value of this projector with respect to the many-body wave function yields the probability $p_j(t)=\braket{\Psi(t)|P_j(t)|\Psi(t)}$ of finding a particle in the $j$th one-body eigenstate.

In order to unravel the nature of the particle transport so as to answer the question of whether it is a tunneling or over-barrier process, it is instructive to subdivide the set of one-body eigenstates into two categories.
First, we introduce the set $\mathcal{B}_{\mathrm{A}}(t)$ that contains all states that lie below the central barrier, i.e., all states $\ket{\Phi_i(t)}$ with eigenenergies $\varepsilon_i(t)<V(x=0,t)$.
Second, $\mathcal{B}_{\mathrm{B}}(t)$ captures all remaining trapped states, i.e., all states $\ket{\Phi_i(t)}$ with eigenenergies $V(x=0, t)\leq\varepsilon_i(t) <0$.
It should be noted that both the eigenenergies and the central potential, and consequently, also the sets $\mathcal{B}_{\sigma}(t)$ change over time.

As the next step we construct the operators
\begin{equation}
  O_{\sigma}(t)=\sum\limits_{\substack{j\text{ such that} \\ \ket{\Phi_j(t)}\in\mathcal{B}_{\sigma}(t)}} P_j(t),\qquad\sigma\in\lbrace\mathrm{A},\mathrm{B}\rbrace
\end{equation}
that project the many-body wave function onto the states in the respective basis sets.
The expectation values $\ev{O_{\sigma}(t)}$ can be understood as the probabilities of a particle to occupy any of the states included in the corresponding basis set $\mathcal{B}_{\sigma}(t)$.
Additionally, we define the operator $O_{\mathrm{C}}(t)=1-O_{\mathrm{A}}(t)-O_{\mathrm{B}}(t)$ that projects the wave function onto the orthogonal space of all untrapped eigenstates.
Consequently, the expectation value $\ev{O_{\mathrm{C}}(t)}$ correctly captures the occupation of the untrapped continuum which is discretized due to our finite numerical grid.

Figure~\ref{fig:spfComponents} shows examples for the time evolution of these quantities.
In the initial state, only under-barrier states are occupied and hence $\ev{O_{\mathrm{A}}(t)}\approx 1$ in the beginning of the time evolution.
As the wells start to penetrate each other during part (II) of the dynamics, the occupation of the under-barrier states $\ev{O_{\mathrm{A}}(t)}$ drops to zero while the occupation $\ev{O_{\mathrm{B}}(t)}$ of the trapped over-barrier states rises to approximately one.
Consequently we classify the particle transport that occurs during this stage of the time evolution as an over-barrier process.
A deeper analysis shows that the start of transport coincides with the crossing of $V(x=0,t)$ of the eigenenergies $\varepsilon_1(t)$ and $\varepsilon_2(t)$ (see Fig.~\ref{fig:oneBodySpectrum}).
The corresponding states $\ket{\Phi_1(t)}$ and $\ket{\Phi_2(t)}$ are predominantly occupied (see Fig.~\ref{fig:spfOccupation}).
Consequently, the particle transport occurs when these states lie above the central barrier.
Towards the end of the propagation, the over-barrier states become under-barrier states again such that $\ev{O_{\mathrm{A}}(t)}\to 1$ while $\ev{O_{\mathrm{B}}}(t)\to 0$ for $t\to t_{\mathrm{f}}$.

For fast collisions [see Figs.\ 6(c) and 6(d)] untrapped states come into play as can be seen in an increase of $\ev{O_{\mathrm{C}}(t)}$ towards the end of the dynamics.
We analyze this phenomenon further in Sec.~\ref{sec:unbinding} where we investigate the emission of particles.
\begin{figure}[htbp]
  \centering
  \includegraphics{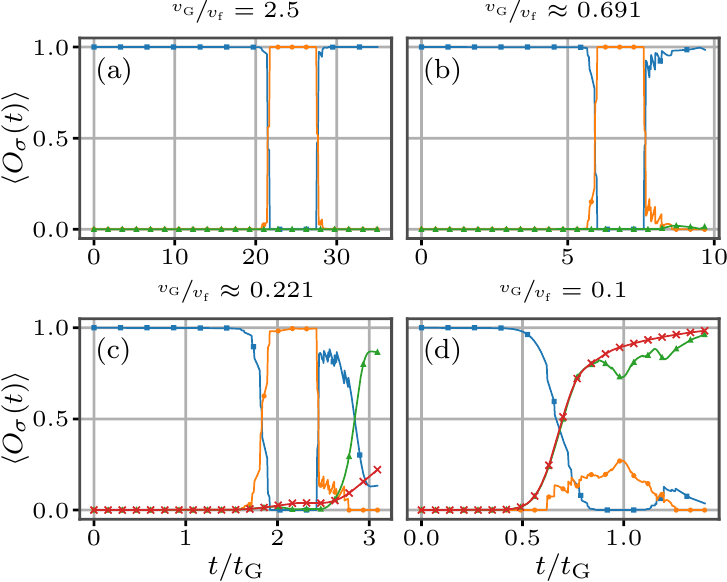}
  \caption{Time evolution of the projections $\ev{O_{\mathrm{A}}(t)}$ (blue solid line with squares), $\ev{O_{\mathrm{B}}(t)}$ (orange solid line with circles), and $\ev{O_{\mathrm{C}}(t)}$ (green solid line with triangles) for different final speeds $v_{\mathrm{f}}^{-1}$. In (c) and (d) we also show the evolution of $\ev{O_{\mathrm{C}}(t)}$ if the initially right well is absent during the propagation ($V_0^{\prime}=0$, red solid lines with crosses) in order to highlight the influence of the second well on the deconfinement of the particles (see Sec.~\ref{sec:unbinding}).}\label{fig:spfComponents}
\end{figure}

\subsection{Deconfinement of particles}\label{sec:unbinding}
As the next step in our analysis, we investigate the origin of the faint density halo between the wells that we observe for fast collisions [see Fig.~\ref{fig:gpop} (e)], indicating a deconfinement of particles.
The increase of $\ev{O_{\mathrm{C}}(t)}>0$ in Fig.~6(c) and 6(d) shows that indeed untrapped delocalized eigenstates of the one-body Hamiltonian $h(x,t)$ [see Eq.~\ref{eq:one_body_hamiltonian}] come into play.
In order to understand how the occupation of the individual eigenstates evolves over time, we analyze the probabilities $p_j(t)=\ev{P_j(t)}$ of finding a particle in a specific one-body eigenstate.
Figures~7(a)--7(d) show the time evolution of these quantities for specific values of $v_{\mathrm{f}}^{-1}$.
\begin{figure}[htbp]
  \centering
  \includegraphics{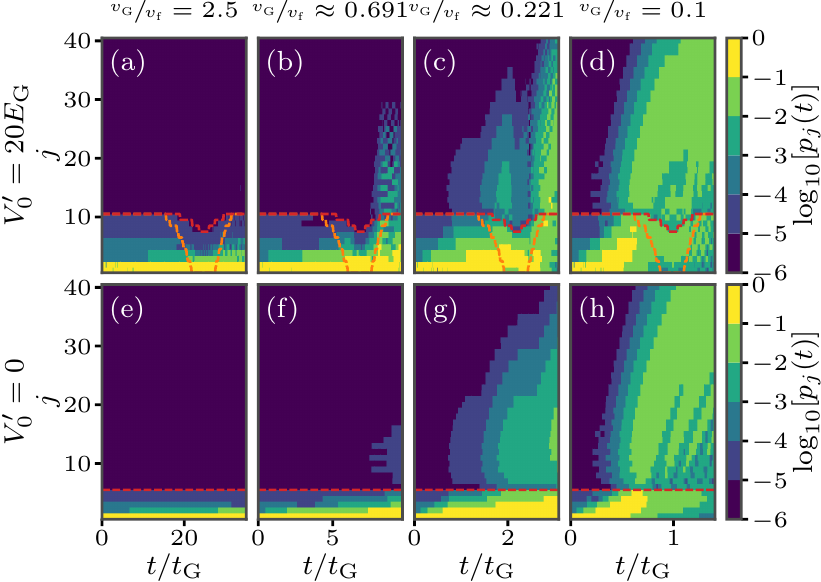}
  \caption{Time evolution of the occupations $\log_{10}[p_j(t)]$ of the $40$ energetically lowest, instantaneous eigenfunctions of the one-body Hamiltonian~\eqref{eq:one_body_hamiltonian}. (a)--(d) the occupation under the presence of the well centered around $\mu^\prime(t)$ and (e)--(h) the case $V_0^\prime=0$. All states below the red dashed line are trapped states, while the states below the orange line are under-barrier states.\label{fig:spfOccupation}}
\end{figure}
For slow collisions [see Fig.~\ref{fig:spfOccupation}(a)] we observe that the eigenstates $\ket{\Phi_1(t)}$ and $\ket{\Phi_2(t)}$ are predominantly occupied while the other excited trapped states play a minor role and no occupation of the untrapped states takes place.
When increasing the acceleration and hence the collision speed, we observe a higher occupation of the excited trapped states and a minor population of several untrapped ones [see Fig.~\ref{fig:spfOccupation}(b)].
For the fastest collisions under consideration [see Figs.~7(c) and 7(d)] all $40$ depicted eigenstates play a significant role and we even observe an equal population of all eigenstates towards the end of the simulation.

We remark that the occupation of untrapped states occurs at different stages of the dynamics when comparing Figs.~7(b)--7(d).
In Fig.~\ref{fig:spfOccupation}(b) the population of untrapped states increases abruptly towards the end of the considered dynamics while still remaining small overall $\ev{O_{\mathrm{B}}(t)}\ll 1$ [see Figure~\ref{fig:spfComponents}(b)].
A similar jump in the occupation of untrapped states towards the end of the dynamics is visible in Fig.~\ref{fig:spfOccupation}(c) albeit with a much stronger total occupation of untrapped states $\ev{O_{\mathrm{C}}(t_{\mathrm{f}})}\approx 0.86\gg \ev{O_{\mathrm{A}}(t_{\mathrm{f}})}+\ev{O_{\mathrm{B}}(t_{\mathrm{f}})}$.
Here we also observe an additional steady increase in the population of untrapped states that already starts in part (I) of the time evolution.
Even though this is a small effect, it still suggests the existence of two distinct mechanisms of the particle deconfinement.
For very fast collisions [see Fig.~\ref{fig:spfOccupation}(d)] the steady increase of the untrapped population becomes dominant.
This enhancement for faster collisions suggests that it is a kinematic effect of the particles which get spilled out of the potential wells due to the fast acceleration.

In order to distinguish between the two effects leading to deconfinement and to unravel their origins, it is instructive to compare the results in Figs.~7(a)--7(d) with simulations where the second, initially empty well is not present, i.e., for $V_0^{\prime}=0$ [see Figs.~7(e)--7(h)].
The first striking difference is the absence of a sudden jump in the occupation of untrapped states towards the end of the time evolution [compare Figs.~7(b) and 7(c) with Figs.~7(f) and 7(g)].
This contribution to the deconfinement can only be explained due to the presence of the second well.
However, the steady increase in the occupation of untrapped one-body states is still present [compare Figs.~7(c) and 7(d) with Figs.~7(g) and 7(h)).
In Fig.~\ref{fig:spfComponents} these observations become even clearer when comparing the evolution of $\ev{O_{\mathrm{C}}(t)}$ with and without the presence of the initially empty well [see Fig.~\ref{fig:spfComponents}].
For very fast collisions [see Fig.~\ref{fig:spfComponents}(d)] the curves match for the biggest part of the dynamics and only deviate slightly towards the end of the time evolution.
Consequently, the presence of the second well plays only a minor role concerning the emission of particles.
For other parameters however [see Fig.~\ref{fig:spfComponents}(c)], the differences are striking and the occupation of untrapped states is greatly enhanced due to the presence of the second well.

As mentioned before, the emission process during early times of the dynamics is of kinematic origin.
We employ the energy of the system as well as its composition to study this phenomenon further.
Figure~\ref{fig:energy} (a) shows the total energy $E(t)$ as a function of $t$ for various inverse final speeds $v_{\mathrm{f}}^{-1}$.
Since we prepare the system in the ground state all energy curves start at the ground state energy $E(t=0)=E_0\approx -33.6E_{\mathrm{G}}$.
When focusing on a very slow motion of the wells (see the curve for $v_{\mathrm{f}}^{-1}v_{\mathrm{G}}=2.5$), the energy remains constant until $t\approx 0.6 t_{\mathrm{f}}$, where it starts to drop as the particles are now impacted by the second potential well.
As the wells separate, the energy increases back to its initial value.
The behavior of the total energy changes gradually as we turn towards faster accelerations.
First, the dip of the energy becomes less deep and a modulation of the energy becomes visible towards the end of the simulated dynamics.
For $v_{\mathrm{f}}^{-1}v_{\mathrm{G}}\approx 0.221$, the total energy exceeds the value zero at the end of the simulations.
Consequently, an emission and untrapping of the particles take place for energetic reasons alone.
As we increase the acceleration further, the total energy exceeds the value zero earlier during the time evolution, e.g., at $t\approx 0.5 t_{\mathrm{f}}$ for $v_{\mathrm{f}}^{-1}v_{\mathrm{G}}\approx 0.221$, and the dip, while the wells are in close proximity, becomes less pronounced.
\begin{figure}[htbp]
  \centering
  \includegraphics{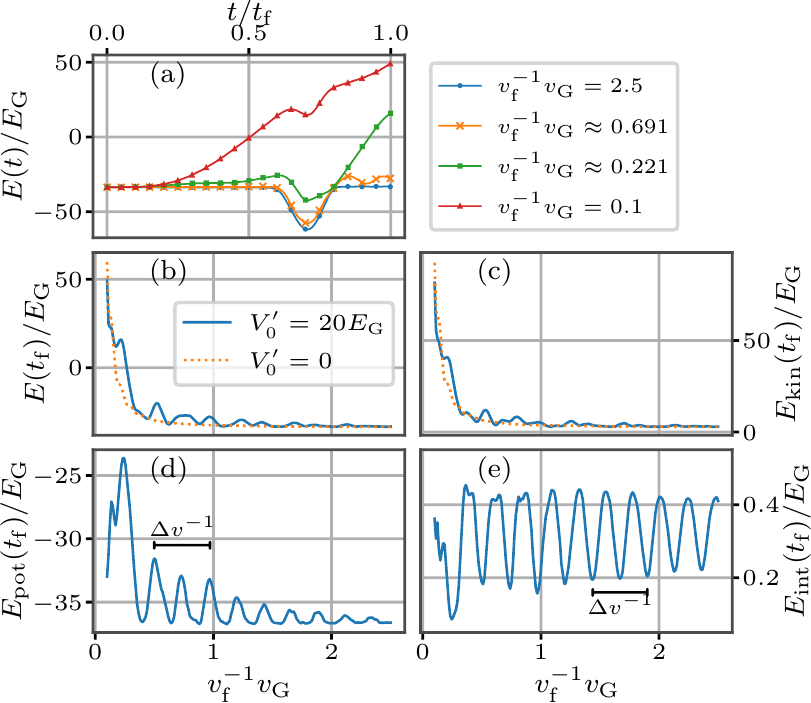}
  \caption{(a) Time evolution of the total energy of the two bosons during the collision dynamics for various inverse final speeds $v_{\mathrm{f}}^{-1}$. Also shown are the (b) total, (c) kinetic, (d) potential, and (e) interaction energies of the final state as a function of $v_{\mathrm{f}}^{-1}$.
    The orange dotted lines in (b) and (c) correspond to computations performed in the absence of the second, initially right, well, i.e. $V_0^\prime=0$, thereby highlighting the impact of this well on the total and kinetic energies.\label{fig:energy}}
\end{figure}
As a next step, we analyze the energy composition of the final state to get an overview of all simulations.
Figures~8(b)--8(d) show the total, kinetic, and potential energies of the final state as a function of the final inverse speed $v_{\mathrm{f}}^{-1}$.
We notice a drastic increase of the kinetic [see Fig.~\ref{fig:energy} (c)] and hence the total energy [see Fig.~\ref{fig:energy}(b)] towards large final speeds, i.e., small $1/v_{\mathrm{f}}$.
For $v_{\mathrm{f}}^{-1}v_{\mathrm{G}}<0.266$ with $V_0^\prime=V_0$ as well as for $v_{\mathrm{f}}^{-1}v_{\mathrm{G}}<0.170$ with $V_0^\prime=0$ the total energy exceeds zero, indicating that untrapping takes place solely for kinetic energy reasons.
The potential energy [see Fig.~\ref{fig:energy} (d)] exhibits equidistant peaks whose height increases towards small values of $v_{\mathrm{f}}^{-1}$ as the particles become less deeply trapped.
As indicated in the figure, the difference between neighboring peaks is equal to half of the period $\Delta v^{-1}=0.47v_{\mathrm{G}}$ that we introduced in our discussion of the final c.m.\ position of the particles.
The same characteristics and effects can be seen for the interaction energy [see Fig.~\ref{fig:energy}(e)].
The maxima of the interaction energy coincide with the extrema of $\ev{X}(t_{\mathrm{f}})$ since the interaction energy is higher when both particles reside in the same well.
The potential energy, on the other hand, becomes maximal where $\ev{X}(t_{\mathrm{f}})$ is zero.
In contrast to the potential energy, the interaction energy does not exhibit a strong increase towards small values of $v_{\mathrm{f}}^{-1}$.
Only a marginal increase in the oscillation amplitude of $E_{\mathrm{int}}(t_{\mathrm{f}})$ is visible as the particles become less deeply trapped and are less strongly localized at the well center.
Due to the local nature of the interaction term, the value of the interaction energy is mainly determined by the delocalization of the particles across both wells and less by how deeply they are trapped.

So far, our discussion of the particle untrapping has relied on the projection onto one-body eigenstates.
We conclude our analysis of this phenomenon using a two-body or in general many-body analysis that relies on projecting the many-body wave function onto number states built from the instantaneous eigenbasis of the one-body Hamiltonian.
Let $\mathcal{N}(t)$ be the time-dependent set of all $N=2$-particle number states that can be constructed from all trapped eigenstates of the instantaneous one-body Hamiltonian.
We then define the magnitude $M_{\mathrm{B}}(t)=\sum_{\ket{\vec{n}}\in\mathcal{N}(t)}{\left|\braket{\vec{n}|\Psi(t)}\right|}^2$, which captures the total overlap of the many-body wave function with the number state basis $\mathcal{N}(t)$.
The maximal possible value of $M_{\mathrm{B}}(t)=1$ indicates that the many-body wave function lies completely in the Hilbert space spanned by the basis $\mathcal{N}(t)$, while a value of zero would indicate that $\ket{\Psi(t)}$ is orthogonal to this space.
Consequently, the quantity $M_{\mathrm{U}}(t)=1-M_{\mathrm{B}}(t)$ can then be used to quantify the untrapped fraction, i.e., the projection of the many-body function onto the orthogonal space of untrapped eigenstates.

Figures~9(a)--9(d) show the time evolution of $M_{\mathrm{U}}(t)$ for different values of $v_{\mathrm{f}}^{-1}$.
For slow to moderately fast collisions [see Figs.~9(a) and 9(b)], no deconfinement of particles is visible in the absence of the second well, i.e.\ for $V_0^\prime=0$.
As discussed previously, only the kinematic emission of particles takes place when only a single well is present.
This process is enhanced by the collisional speed and we only observe untrapping for the fastest collisions under consideration [see Figs.~9(c) and 9(d)].
When comparing these results with the simulations with $V_0^\prime=V_0$, the importance of the presence of both wells becomes evident.
For certain values of $v_{\mathrm{f}}^{-1}$ a drastic increase in the untrapped fraction is noticeable that stems from the final stage of the dynamics [see Figs~9(a) and 9(c)].
At very high speeds however, the kinematic untrapping is the dominant contribution to the emission of particles such that the two curves for $M_{\mathrm{U}}(t)$ (single- and two-well dynamics) match each other.

The logarithmic representation of the one-body density in Figs.~9(e)--9(h) increases the visibility of the density halo outside of the wells in contrast to the earlier discussion (see Fig.~\ref{fig:gpop}).
For very fast collisions [see Figure~\ref{fig:fixedNS}(h)], we notice a density halo on the left side of the initially occupied well due to a fraction of the density getting spilled out of the potential wells due to the inertia of the particles.
Furthermore, we observe that in the case of the resonant emission of particles at certain values of $v_{\mathrm{f}}^{-1}$, the density halo is located in the space between the two well trajectories [see Figs.~9(e) and 9(g)].
At other values, where almost no deconfinement takes place, this halo is vanishingly small [see Fig.~\ref{fig:fixedNS}(f)].

\begin{figure}[htbp]
  \centering
  \includegraphics{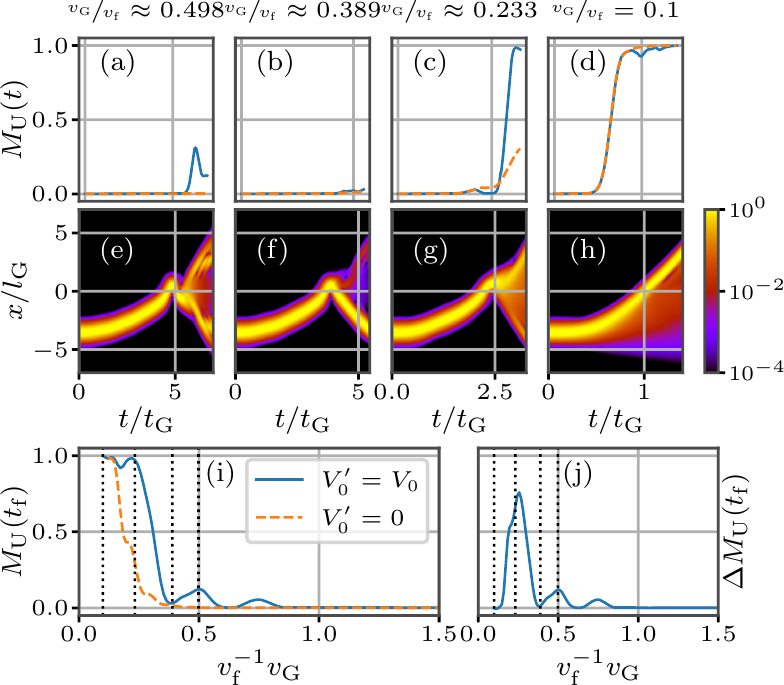}
  \caption{(a)--(d) Time evolution of the untrapped fraction $M_{\mathrm{U}}(t)$ for varying $v_{\mathrm{f}}^{-1}$ (blue solid lines). The orange dashed lines indicate the evolution of $M_{\mathrm{U}}(t)$ in the absence of the second, initially empty well (i.e., $V_0^\prime=0$), highlighting its importance for the untrapping process for certain values of $v_{\mathrm{f}}^{-1}$. (e)--(h) Time evolution of the one-body density $\log_{10}[\rho^{(1)}(x,t)]$ [see Equation~\eqref{eq:gpop}] for $V_0=V_0^\prime$ in a logarithmic representation which increases the visibility of the density halo outside the potential wells in comparison to Fig.~\ref{fig:gpop}. (i) Untrapped magnitude $M_{\mathrm{U}}(t_{\mathrm{f}})$ of the final state as a function of $v_{\mathrm{f}}^{-1}$. The dotted vertical lines indicate the values of $v_{\mathrm{f}}^{-1}$ that have been used for (a)--(d) and (e)--(h). (j) Untrapped magnitude $\Delta M_{\mathrm{U}}(t_{\mathrm{f}})$ due to the presence of the second well (see the main text for details).}\label{fig:fixedNS}
\end{figure}
Figure~\ref{fig:fixedNS}(i) shows the value of $M_{\mathrm{U}}(t)$ for the final state.
In the absence of the second well, i.e.\ for $V_0^\prime =0$, the curve of $M_{\mathrm{U}}(t_{\mathrm{f}})$ is flat and close to a value of zero for $v_{\mathrm{f}}^{-1}v_{\mathrm{G}}\gtrapprox 0.39$ since only the kinematic emission of particles can occur which requires high speeds.
When exceeding this threshold for the final speed, the untrapped fraction rapidly grows and reaches the maximal possible value of one.
In the presence of the second well ($V_0^\prime=V_0$), $M_{\mathrm{U}}(t_{\mathrm{f}})$ exhibits peaks in the parameter regime $v_{\mathrm{f}}^{-1}v_{\mathrm{G}}\gtrapprox 0.39$ that are not present for $V_0^\prime =0$.
Figure~\ref{fig:fixedNS}(j) shows the difference $\Delta M_{\mathrm{U}}(t_{\mathrm{f}})$ between the simulations with $V_0^\prime=V_0$ and $V_0^\prime =0$.
This removes all contributions to the untrapping process that exclusively stem from the acceleration and not from the influence of the second well.
We are able to identify three distinct peaks at $0.257v_{\mathrm{G}}^{-1}$, $0.498v_{\mathrm{G}}^{-1}$, and $0.751v_{\mathrm{G}}^{-1}$ where the emission of particles is resonantly enhanced.
The difference $\Delta M_{\mathrm{U}}(t_{\mathrm{f}})$ as a function of $v_{\mathrm{f}}^{-1}$ is reminiscent of an ionization spectrum.

\subsection{Interparticle correlations and entanglement}
We now analyze the emergence of correlations and entanglement during the collision dynamics by employing the von Neumann entropy~\cite{neumann1927}, which reads
\begin{equation}
  \label{eq:von_neumann}
  S^{(1)}(t)=-\Tr\left\{\hat{\rho}^{(1)}(t)\ln\left[\hat{\rho}^{(1)}(t)\right]\right\}=-\sum\limits_{i=1}^{M}\lambda_i(t)\ln[\lambda_i(t)].
\end{equation}
Here $\hat{\rho}^{(1)}(t)$ refers to the one-body density matrix~\cite{sakmann2008} with eigenvalues $\lambda_i(t)$.
It should be noted that the natural populations $\lambda_i(t)$ possess the property $0\leq\lambda_i(t)\leq 1$ and fulfill the relation $\sum_{i=1}^M\lambda_i(t)=1$.

A value of $S^{(1)}(t)=0$ indicates a mean-field state and implies the absence of any correlations between the two particles.
In the same light, a finite value of $S^{(1)}(t)\neq 0$ corresponds to interparticle correlations and hence a deviation from the mean-field product state.
For a maximally entangled state within our simulations using six SPFs, the von Neumann entropy reaches the maximal value of
\begin{equation}
  \label{eq:von_neumann_max}
  S_{\max}^{(1)}=\ln(M)=\ln(6)\approx 1.79
\end{equation}
which is here solely determined by the dimensionality of the one-body Hilbert space $M=6$.

Figure~\ref{fig:entropy} shows the entropy of the final state as a function of the final inverse speed normalized to the maximal possible value.
We observe a structure of equidistant peaks of varying height indicating large values of $S^{(1)}(t_{\mathrm{f}})$.
The spacing is approximately equal to the period $\Delta v^{-1}=0.47 v_{\mathrm{G}}^{-1}$ obtained during the c.m.\ analysis, suggesting a relation to the final location of the particles.
This hypothesis can be easily confirmed by analyzing the one-body density and the c.m.\ observable, which show that the maxima of the von Neumann entropy correspond to situations where the particles are distributed uniformly over both wells in the final state.
Furthermore, we notice that the entropy reaches its largest value of $S^{(1)}(t_{\mathrm{f}})\approx 0.715 S_{\max}$ for $v_{\mathrm{f}}^{-1}v_{\mathrm{G}}\approx 1.21$, indicating a highly entangled state for which the two largest natural populations are almost equal [$\lambda_1(t_{\mathrm{f}})\approx 0.517$ and $\lambda_2(t_{\mathrm{f}})\approx 0.479$].
The minima between the peaks correspond to values of $v_{\mathrm{f}}^{-1}$ where the particles are localized in one of the wells, i.e., extrema of the c.m.\ position.
Here the first natural population is dominant $\lambda_1(t_{\mathrm{f}})\approx 1$.
\begin{figure}[htbp]
  \centering
  \includegraphics{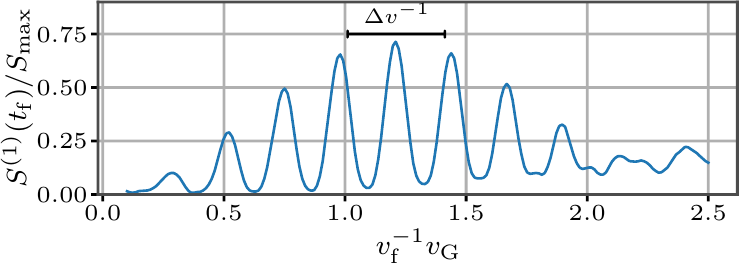}
  \caption{von Neumann entropy of the final state $S^{(1)}(t_{\mathrm{f}})$ normalized by the maximal possible value $S_{\max}^{(1)}$ as a function of the inverse final speed $v_{\mathrm{f}}^{-1}$.\label{fig:entropy}}
\end{figure}
We notice that the height of the local maxima decreases towards faster collisions and the entropy drops to zero, indicating a mean-field product state.
The reason for this behavior is that for $v_{\mathrm{f}}^{-1}\to 0$ the first natural population becomes dominant $\lambda_1\approx 1$.
When considering slow collisions ($v_{\mathrm{f}}^{-1}v_{\mathrm{G}}\gtrapprox 2$), the peak structure of $S^{(1)}(t_{\mathrm{f}})$ vanishes but the entropy does not drop to zero.
This indicates that still measurable correlations between the two particles exist.

\section{Conclusions and outlook}\label{sec:conclusions}
We have investigated the collisional nonequilibrium quantum dynamics of ultracold bosons confined in two colliding potential wells.
We were able to subdivide the dynamics into three distinct stages by identifying the underlying physical processes.
Initially, the particles follow the trajectories of the wells closely.
When the well separation falls below a certain threshold, a periodic collective particle transport takes place in an effective time-dependent double-well structure.
By analyzing the population of SPF states we were able to classify this transport as an over-barrier process.
Using the c.m.\ position of particles, we have been able to quantify the number of oscillatory transitions that occur during the dynamics.
During the separation of the wells in the third part of the time evolution, we noticed a mode motion of the particles within each well.
The amplitude of this motion depends on the location of the particles with respect to the well centers at the end of the collision process.
We determined a phase of $\sfrac{\pi}{2}$ between the dipole modes of both wells while the frequency of this motion is independent of the acceleration.
Furthermore, we observed that for certain final speeds the particles are strongly localized in one of the wells while they are generally delocalized.
This phenomenon resembles the charge transfer that takes place during atom-atom collisions.
Another important feature of our time-dependent setup is the untrapping of particles which we characterized in detail using a SPF, number state, and energetic analysis.
We have been able to quantify the untrapped fraction unraveling two different contributions to it.
During fast collisions, the kinetic energy grows continuously, which leads to a positive total energy and consequently to a particle untrapping.
However, we also observed a resonant untrapping effect for certain kinematic parameters leading to a rapid emission of particles as the wells separate.
We have been able to determine the dependence of this second mechanism on the kinematic parameters, which is reminiscent of an ionization spectrum.

Our findings serve as a promising starting point for further studies in different directions.
By increasing the interparticle interaction strength one could enhance the amount of correlations that arises during the dynamics and it would be interesting to explore the corresponding impact on the resonant particle untrapping.
A variation of the potential wells, for example, by decreasing the depth or introducing an asymmetry between the two Gaussians could modify the particle transport.
In this context, a more detailed study of the correlation and the creation of entanglement, incorporating the spatial and momentum space resolution of correlation functions, might be instructive~\cite{bergschneider2019,becher2020}.
In the light of atom-atom collisions, a particularly intriguing prospect is to employ different initial states.
Employing an initial state that incorporates particles in both wells could lead to an enhancement of the emission due to opposite momenta of the bosons.
Furthermore, it would be interesting to investigate the impact of the trajectories of the wells.
Finally, the multiconfiguration time-dependent Hartree method for fermions~\cite{zanghellini2003,caillat2005} allows one to study the nonequilibrium dynamics of fermions in a similar setup.
It would be instructive to analyze the role of the particle statistics and how the phenomena described in this work might be modified.

Another exciting route would be the investigation of mixtures of different components which is of particular interest for ultracold-atom research.
Such ensembles can be composed of different elements~\cite{ferrari2002,brue2012}, isotopes~\cite{poli2005} or hyperfine states~\cite{myatt1997} and exhibit a plethora of exciting and unique properties such as relative phase evolution~\cite{anderson2009}, composite fermionization~\cite{garcia-march2013}, nonlinear~\cite{kamchatnov2014} and collective excitations~\cite{mertes2007} as well as miscible-immiscible phase transitions~\cite{ticknor2013,nicklas2015}.
Depending on the particle statistics this allows for the realization of Bose-Bose~\cite{modugno2002,catani2008}, Fermi-Fermi~\cite{wille2008,kohstall2012} and Bose-Fermi mixtures~\cite{schreck2001,hadzibabic2002,ospelkaus2006,heinze2011}.
The multilayer multiconfiguration time-dependent Hartree method for mixtures~\cite{cao2017} is a powerful numerical approach to treat the correlated nonequilibrium dynamics of such systems which allows one to extend the setup presented in the present work to such mixtures.
The role of the interspecies interaction as well as a possible massimbalance between the constituents are particularly of interest.

\begin{acknowledgments}
  The authors acknowledge fruitful discussions with K.\ Keiler.
  This work was funded by the Deutsche Forschungsgemeinschaft (German Research Foundation) SFB 925 Project No.\ 170620586.
\end{acknowledgments}

\appendix
\section{Technical aspects and convergence}\label{sec:convergence}
In the present work we employ the fast Fourier transform (FFT)~\cite{kosloff1983,kosloff1988,beck2000} to obtain a spatially discretized representation of the operators and the SPFs.
This scheme allows the efficient numerical treatment of large grids consisting of $n\gtrapprox 100$ grid points compared to another approaches relying on discrete variable representations (DVRs)~\cite{beck2000}.
We use $n=675$ grid points that are equally spaced in the interval $\left(-7l_{\mathrm{G}},7l_{\mathrm{G}}\right]$.
It should be noted that the FFT scheme implies periodic boundary conditions for the physical system.
We repeat the same set of simulations presented in the main text using a sine DVR~\cite{beck2000} which incorporates hard-wall boundary conditions.
Thereby we are able to confirm that spacing between the potential wells and the edges of the grid is large enough such that no influence of the boundary conditions is visible in the observables discussed in the present work.

The underlying time-dependent variational principle used to derive the MCTDHB equations of motion guarantees that the SPF basis is rotated such that the many-body wave function optimally captures the state of the physical system.
However, care has to be taken in order to ensure that the number $M$ of SPFs is sufficiently large and thereby the numerical convergence of the method is guaranteed~\cite{beck2000,alon2008}.
We compare the results presented in the main text with simulations that include an additional, seventh SPF and observe that the observables discussed in the main text do not change significantly.
The ground state energy exhibits a relative change of the order of $10^{-5}$ and the energy of the final state of $10^{-4}$ in the worst case.
We observe that the untrapped fraction of the final state $\Delta M_{\mathrm{U}}(t_{\mathrm{f}})$ determined changes at most by an absolute value of $4\times 10^{-4}$ when including the additional orbital.
The absolute change in the relative entropy $\sfrac{S^{(1)}}{S^{(1)}_{\mathrm{max}}}$ of the final state is limited by $0.03$.
The center-of-mass position of the particles at the end of the time evolution changes at most by $1\%$.

Additionally, the spectral representation of the one-body density matrix is important to judge the convergence of the approach.
The eigenvalues of $\rho^{(1)}(t)$, the so-called natural populations, should exhibit a rapidly decreasing hierarchy.
This indicates that any natural orbitals (eigenstates of the one-body density matrix) that are neglected due to the truncation of the single particle Hilbert space play a negligible role.
We find that this is the case for all parameters considered in the present work and that the least occupied orbital taken into account shows a population of $\lambda_6<10^{-4}$ for all simulations.
Therefore, we consider $M=6$ SPFs sufficient to describe the time evolution of the physical system accurately.

\end{document}